# Self-aligned multi-channel superconducting nanowire avalanche photodetector

Risheng Cheng, Xiang Guo, Xiaosong Ma, Linran Fan, King Y. Fong, Menno Poot, and Hong X. Tang[a]

*Department of Electrical Engineering, Yale University, 15 Prospect Street, New Haven, Connecticut 06511, USA*

We describe a micromachining process to allow the coupling of an array of single-mode telecommunication fibers to individual superconducting nanowire single photon detectors (SNSPDs). As proof of principle, we show the integration of four detectors on the same silicon chip, including two standard single-section nanowire detectors and two superconducting nanowire avalanche photodetectors (SNAPs) with modified series structure without external inductor, and their performances are compared. The SNAP shows saturated system detection efficiency of 16% while the dark count rate is less than 20 Hz, without the use of photon-recycling reflectors. The SNAP also demonstrates doubled signal-to-noise ratio, reduced reset time (~ 4.9 ns decay time) and improved timing jitter (62 ps FWHM) compared to standard SNSPDs.

In recent years, superconducting nanowire single-photon detectors (SNSPDs)[1,2] have emerged as successful alternatives to traditional InGaAs/InP-based single-photon avalanche photodiodes (SPADs) in the realm of near-infrared single photon detection due to their excellent quantum efficiency,[3–6] short timing jitter,[6–8] ultralow dark count rates,[9–11] fast reset time with only several nanoseconds,[6,12,13] and photon number resolving ability.[14–16] To date, most of SNSPD systems utilize fiber coupling method because of advantages over free-space coupling, including lower dark count rates, more compact size and robustness to vibration. However, precise alignment between the beam-spot (~10 μm in diameter) from single-mode fiber and active nanowire area (typically 10-15 μm in diameter) still remains challenging, in particular for robust, multi-cycle operation at low temperatures. Three dimensional cryogenic positioner in combination with fiber focuser is commonly employed for *in situ* matching of the beam waist to the detection area of the nanowire detector.[17] High-efficiency fiber-to-detector coupling can be realized in this way but only one detector can be addressed at a time. Recently, a self-aligned coupling method based on deep-etching of silicon wafer was proposed by the NIST group.[18] Using this technique, a front-illuminated WSi-based SNSPD was realized and achieved an impressive 93% system efficiency, boosted by a reflector and cavity embedded between the superconducting nanowire and the substrate.[4] Here, we demonstrate a backside silicon micromachining process that allows the precision placement of cleaved fiber in the etched pit in a self-aligned fashion while NbTiN superconducting nanowires are fabricated on the front side.

[a] Electronic mail: hong.tang@yale.edu

Multiple fibers are installed in an array of patterned pits on a single detector chip. This approach enables back-illuminated detector structure which separates the optical access and electronic readout circuits on two sides of the chip and allows for compact integration of multi-channel detectors.

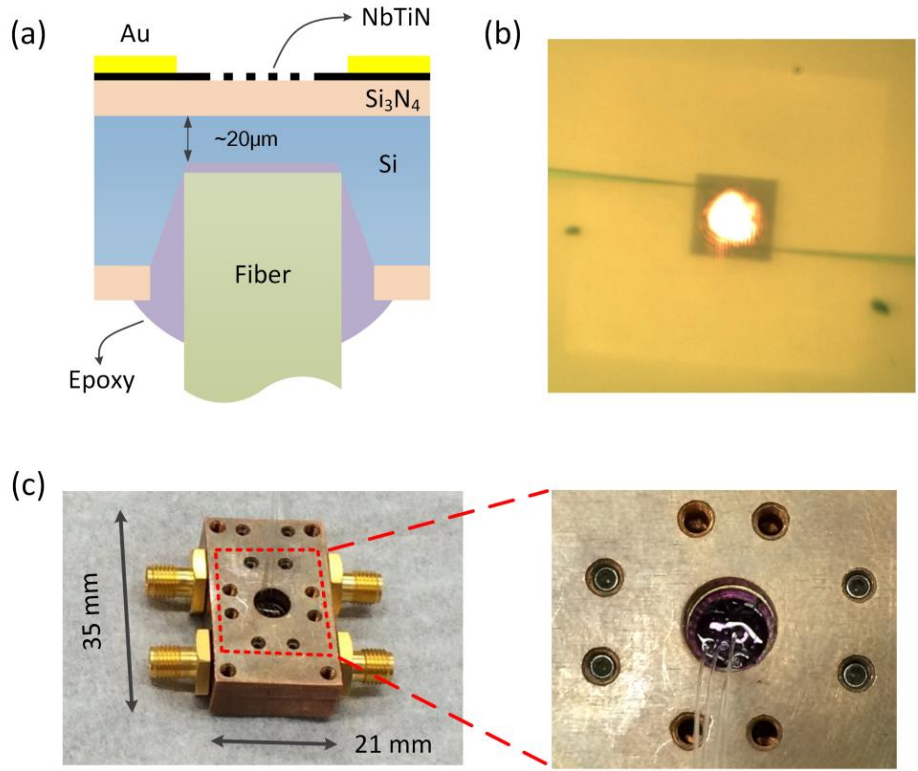

FIG. 1. (a) Cross-sectional sketch of the nanowire detector with self-alignment structure (not to scale). (b) Top view optical micrograph of detector area with a laser spot illuminated from backside fiber. The light rectangular is $Si_3N_4$/Si membrane and the nanowire is at the center of the membrane. (c) Photographs of fiber-coupled detector package. Four fibers are connectorized with detectors on a single chip.

Figure 1(a) presents a cross-sectional view of our detector structure. The fabrication begins with a double-side polished silicon wafer, on both sides of which 200 nm-thick LPCVD silicon nitride ($Si_3N_4$) is deposited. Then, arrays of rectangular-shaped windows are opened in the nitride layer at the backside via optical lithography and following reactive ion etching (RIE). Using these nitride patterns as mask, the silicon substrate can be anisotropically etched into trapezoidal pits in heated KOH solution. In principle, the etching stops automatically when the silicon substrate is completely etched through, leaving free-standing nitride membranes. However, we stop the reaction in advance by accurate timing and thus leaving 20 μm -thick residue silicon layer, which serves as supporting structure to enhance the mechanical strength of the membranes. By controlling the dimension of the windows patterned in the nitride mask layer, we are able to make the bottom of silicon pits slightly smaller than a standard SMF-28 fibers (125 μm in diameter) so that fibers can be stuck at certain height without breaking the membrane as shown in Fig. 1(a). After the wet-etching process, a thin layer of NbTiN superconducting film is deposited by means of DC magnetron sputtering and the silicon wafer is then diced into several smaller chips (8 mm × 11 mm size), which contain an array of 4 × 5 = 20 pits, for further device fabrication. A total of 20 nanowire detectors are fabricated on each chip via two e-beam lithography steps. The first e-beam lithography exposes the nanowire pattern in 6% HSQ (~100nm thickness) negative resist, which is accurately aligned with the backside silicon pits so that the active nanowire area of the detector is centered in the nitride membrane with a misalignment less than 1 μm. The second e-beam lithography defines Ti/Au metal pads using a standard PMMA/MMA bilayer lift-off process. Finally, the nanowire pattern of HSQ is transferred to underlying NbTiN layer by RIE using $CF_4$ plasma.

The completed detector chip is mounted on a printed circuit board (PCB) and fixed upside down on an inverted microscope for fiber alignment and packaging. The targeted pit is first filled with a tiny drop of UV-curable epoxy. A cleaved fiber is manipulated by a 3-axis motorized stage with a minimum step size of 30nm to approach the bottom of the pit. Precision alignment to the active nanowire area is guided by a red laser spot launched through the fiber as shown in Fig. 1(b). The epoxy is cured by exposing the chip using a UV gun for 1-2 minutes. Likewise, three other fibers are aligned and glued to different detectors. The four chosen nanowires are wire-bonded to the coplanar waveguides on the PCB which is in turn installed in a copper box shown in Fig. 1(c). As the last step, the backside of the detector chip is flooded with the UV epoxy and cured. This final step is important for achieving a very robust package of multi-channel detectors. In principle, one could wire up all the 20 nanowire detectors fabricated in one chip but we are currently limited to four detectors to match the number of available coaxial cables in the cryostat.

Among the total 20 nanowire detectors fabricated on one silicon chip, 10 detectors are standard SNSPDs and the other 10 are superconducting nanowire avalanche photodetectors (SNAPs). Different from the $n$-SNAPs reported by Ref.[20] and Ref.[21], which employ current-limiting inductors serially connected with $n$ parallel active nanowires, we adopted the modified structure of series-2-SNAP similar to Ref [22] but with much larger detection area. In this way, the external inductor could be entirely folded into the active detection area and thus the total inductance of the detector can be significantly reduced. Two standard SNSPDs and two series-2-SNAPs were randomly selected among the 20 detectors for fiber connectorization.

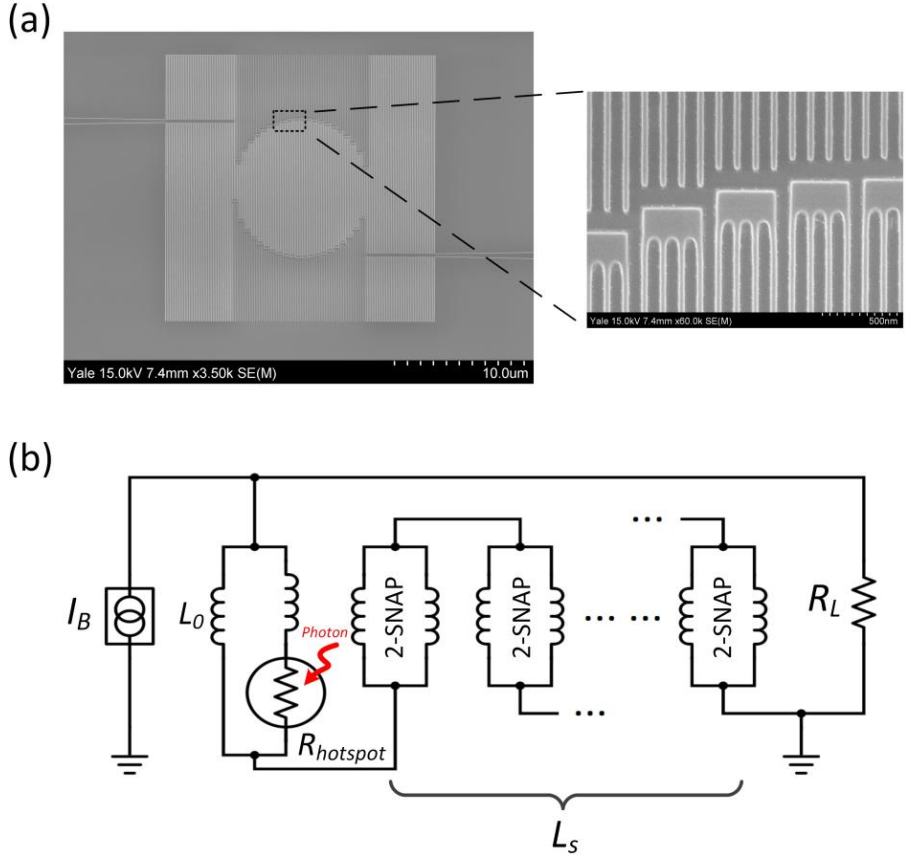

FIG. 2. (a) Scanning electron micrograph of a series-2-SNAP and higher magnification image taken at the corner of the active nanowire area. The diameter of detection area is 10 μm. (b) Equivalent electrical circuit diagram for series-2-SNAP, where every nanowire section is model as an inductor $L_0$. $I_B$ represents bias current and $R_L$ the load resistance. In this modified structure, all the unfired nanowire pairs serve as current-limiting series inductor $L_s$ until the avalanche happens.

Figure 2(a) show scanning electron micrographs (SEMs) of our series-2-SNAP. The thickness of NbTiN film deposited on $Si_3N_4$-on-Si substrate is 6.5nm covered by 2nm-thick native oxide (confirmed by TEM). The nanowire width, spacing and the diameter of the circular-shaped active nanowire area is 40nm, 80nm and 10 μm, respectively. The floating nanowires outside the circular area is for proximity effect correction during e-beam exposure, which constitutes 20 μm × 20 μm rectangular together with the active nanowires. The 180 degree bending between nanowire elements is especially designed as round corner for relieving current crowding.[23] Figure 2 (b) shows an equivalent electrical circuit model for series-2-SNAP. The basic 2-SNAP element consists of two parallel nanowires which are modeled as two inductors ($L_0$), and the series-2-SNAP consists of an array of serially connected 2-SNAPs. When a single photon is absorbed by any nanowire section and thus create a hotspot, the current has to be diverted into the parallel nanowire first because of the current-limiting serial inductor $L_s$. If the bias current $I_B$ is high enough (larger than avalanche current $I_{AV}$), the diverted current will switch the secondary nanowire section to normal state. Therefore, most of the current flowing through the device, which is about two time the current carried by a single nanowire section, will be diverted to the read-out $R_L$, providing two times signal-to-noise ratio (SNR) compared to standard SNSPDs. In previous *n*-SNAP structure,[20,21] the current-limiting series inductor $L_s$ is connected externally and typically designed as 10 times larger than $L_0$ for ensuring stable operation without afterpulsing,[24] which limits the reset time and also timing jitter. However, in this modified series-2-SNAP structure, all the unfired nanowire pairs serve as current-limiting inductor $L_s$ until the avalanche happens. If the active area of the detector is large, e.g. 10 μm in diameter, the total length of the nanowire is more than 650 μm. This is equivalent to $L_s > 16 \times L_0$, so that dedicated external inductor is no longer needed and hence the reset time can be shortened considerably.

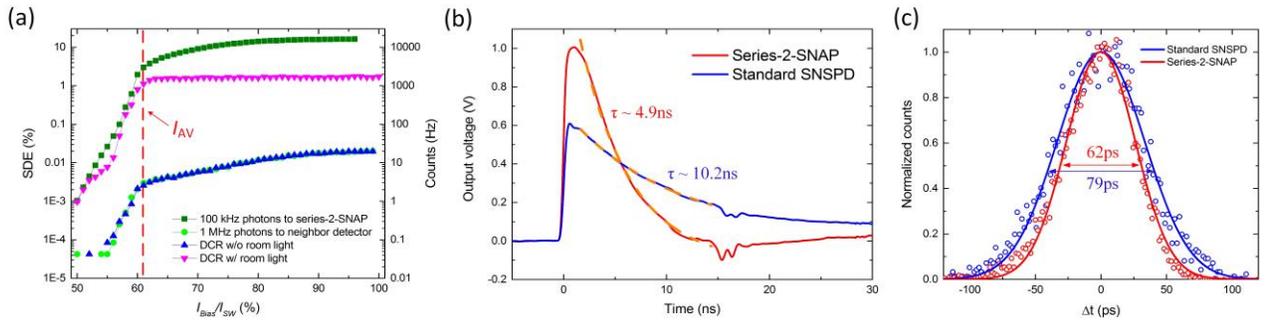

FIG. 3. (a) SDE and DCR as a function of normalized bias current $I_{Bias}/I_{SW}$. The switching current $I_{SW}$ is measured to be 18.6 μA. (b) Averaged output pulses from series-2-SNAP and standard SNSPD. The decay time constants are extracted from exponential fitting. (c) Histogram data obtained from the jitter measurement (empty circles) and corresponding Gaussian fit (solid line). The double-arrows indicate full width at half maximum (FWHM) obtained from the fits. The series-2-SNAP and standard SNSPD are biased at 16 μA and 8 μA, respectively.

In order to characterize the performance of our detectors, the detector package is mounted on the cold plate of a closed-cycle refrigeration cryostat[25] and cooled down to ~1.7K. 1550nm laser light is sent through single-mode fibers installed in the cryostat, which are fusion-spliced with the fibers in the detector package. Polarization controller is used for optimizing the polarization status of incident photons and the photon flux is fixed at 100 kHz via a series of variable attenuators. Figure 3 (a) shows the plot of the system detection efficiency (SDE) and dark count rate (DCR) as a function of normalized bias current for the best detector (series-2-SNAP). We measure the maximum SDE of 16.1% for parallel polarized photons with the DCR (without room light) lower than 20 Hz. However, we find the DCR increases to ~1.6 kHz with room light switched on, which we believe arises from the stray light that leak into fibers and cryostat and can be removed by better shielding of fibers and cryostat. In order to investigate the crosstalk between neighboring detectors we sent 1 MHz rates of photons to the adjacent detector and measured the dark count again, the curve of which perfectly overlaps with the DCR curve without room light, indicating the absence of crosstalk. Figure 3(b) and (c) show averaged signal traces and timing jitter measured by oscilloscope for series-2-SNAP and standard SNSPD having the same detection area, respectively. As expected, the SNR of series-2-SNAP is almost doubled compared with standard SNSPD and the decay time is shortened from 10.2 ns to 4.9 ns, which are extracted from exponential fitting. Series-2-SNAP also shows better time performance with reduced jitter of 62 ps compared to standard SNSPD's 79 ps owing to the improved SNR. We also measured the performance of the other three detectors and they demonstrate 12% (series-2-SNAP) and 3-5% (standard SNSPDs) system efficiency at similar dark count level.

In conclusion, we demonstrated a new robust self-aligned packaging scheme with multi-channel detectors on a single chip. We also showed saturated SDE for the first time, to the best of our knowledge, in series-2-SNAP with detection area comparable with the core size of single-mode fibers. We demonstrate good success with all four connected detectors. In principle, larger array of detectors can be assembled if sufficient coaxial readout cables are available. Furthermore, by integrating the detectors with optical cavities combined with pre-screening before packaging, it is possible to realize more efficient multi-channel single-photon detectors in a very compact size. Recently, our multi-channel detector package has been already used in characterizing fiber-coupled photon-pair sources.

This work was supported by Packard Foundation and DARPA. The authors thank Michael Power, James Agresta, Christopher Tillinghast and Dr. Michael Rooks for the assist provided in device fabrication.